\documentclass[aps,prl,twocolumn,superscriptaddress,amsmath,amssymb]{revtex4-1}

\usepackage{amsmath}
\usepackage{graphicx}
\usepackage{amsfonts}
\usepackage{amssymb}
\usepackage{color}
\usepackage{dcolumn}% Align table columns on decimal point
\usepackage{bm}% bold math

\newcommand{\be}{\begin{equation}}
\newcommand{\ee}{\end{equation}}
\newcommand{\bea}{\begin{eqnarray}}
\newcommand{\eea}{\end{eqnarray}}

\begin{document}

%\preprint{APS/123-QED}

\title{Field-induced superdiffusion and dynamical heterogeneity}

\author{Giacomo Gradenigo}
\affiliation{Universit\'e Grenoble Alpes, LIPHY, F-38000 Grenoble, France \\ CNRS, LIPHY, F-38000 Grenoble, France}

\author{Eric Bertin}
\affiliation{Universit\'e Grenoble Alpes, LIPHY, F-38000 Grenoble, France \\ CNRS, LIPHY, F-38000 Grenoble, France}

\author{Giulio Biroli}
\affiliation{IPhT, Universit\'e Paris Saclay, CEA, CNRS, F-91191 Gif-sur-Yvette Cedex, France \\ Laboratoire de Physique Statistique, Ecole Normale Sup\'erieure, PSL Research University; Universit\'e Paris Diderot Sorbonne Paris-Cit\'e; Sorbonne Universit\'es UPMC Univ. Paris 06; CNRS; 24 rue Lhomond, 75005 Paris, France. }

\date{\today}% It is always \today, today,
             %  but any date may be explicitly specified

\begin{abstract}
 
  By analyzing two Kinetically Constrained Models of supercooled
  liquids we show that the anomalous transport of a driven tracer
  observed in supercooled liquids is another facet of the phenomenon
  of dynamical heterogeneity. We focus on the Fredrickson-Andersen and
  the Bertin-Bouchaud-Lequeux models.  By numerical simulations and
  analytical arguments we demonstrate that the violation of the
  Stokes-Einstein relation and the observed field-induced
  superdiffusion have the same physical origin: while a fraction of
  probes do not move, others jump repeatedly because they are close to
  local mobile regions.  The anomalous fluctuations observed out of
  equilibrium in presence of a pulling force $\epsilon$,
  $\sigma_x^2(t) = \langle x_\epsilon^2(t) \rangle - \langle
  x_\epsilon(t) \rangle^2 \sim t^{3/2}$, which are accompanied by the
  asymptotic decay $\alpha_\epsilon(t)\sim t^{-1/2}$ of the non-Gaussian
  parameter from non-trivial values to zero, are due to the splitting
  of the probes population in the two (mobile and immobile) groups and
  to dynamical correlations, a mechanism expected to happen
  generically in supercooled liquids.

\end{abstract}

\pacs{Valid PACS appear here}% PACS, the Physics and Astronomy
                             % Classification Scheme.

\maketitle

Superdiffusion in presence of an external driving is one among the
most intriguing results of microrheological numerical experiments in
supercooled liquid~\cite{WHVB12,SH13} and of experimental studies on
glassy granular media~\cite{LCDBB10}. In supercooled liquids the mean
square displacement (MSD) of a tagged particle displays a
characteristic intermediate plateau of increasing length when the
temperature is lowered, due to caging, while asymptotically the
diffusion is always Fickian, $\langle x^2(t) \rangle \sim t$. The
surprising finding of~\cite{WHVB12,SH13} is that, notwithstanding the
slowing down of the dynamics, at low enough temperatures the action of
an external force on a probe particle produces a superdiffusive
spreading of the probability distribution of displacements, namely $
\sigma_x^2(t) = \langle x_\epsilon^2(t) \rangle - \langle
x_\epsilon(t)\rangle^2 \sim t^\gamma$ with $\gamma>1$ and with
$\epsilon$ representing a force acting on the probe. This
\emph{anomalous} behaviour of the mean square displacement (MSD)
around the drift is really a landmark of non-Fickian diffusion: in the
case of Fickian diffusion the MSD around the drift would grow linearly
in time, $\sigma_x^2(t) \sim t $, as the unbiased MSD does.  What is
the mechanism which triggers a faster-than-Fickian diffusion within an
environment with large and broadly distributed relaxation times? At
first sight this is a quite counter-intuitive behaviour. The goal of
this paper is to provide a clear-cut explanation of such a mechanism,
clarifying how this anomalous diffusion is intimately related to
dynamical heterogeneities and to the breaking of the Stokes-Einstein
relation.\\ The Stokes-Einstein (S-E) relation connects the diffusion
coefficient $D(\beta)$ of a probe to the relaxation time
$\tau_{\textrm{eq}}(\beta)$ of the sample as
$D(\beta)\tau_{\textrm{eq}}(\beta)=\textrm{const} $ \cite{foot1}. The
physical reason for such relation is that simple liquids are
characterized by a single relevant time-scale $\tau_D$: this
time-scale characterizes the behaviour of the system at all scales,
from the single molecule diffusion, $D(\beta)\sim\tau_D^{-1}(\beta)$,
to the relaxation of the sample $\tau_{\textrm{eq}}(\beta)\sim
\tau_D(\beta)$, which explains the Stokes-Einstein relation.  Close to
the glass-transition temperature $T_g$ several time-scales appear in
the dynamics of the system. In this regime the characteristic
diffusion time $\tau_D(\beta) \sim D^{-1}(\beta)$ \emph{decouples}
from the relaxational time, $\tau_{\textrm{eq}}(\beta)\nsim
\tau_D(\beta)$, so that one also finds $ D(\beta) \tau_{\textrm{eq}}(\beta) \neq
\textrm{const}$~\cite{kivelsontarjusonSE}: this is the Stokes-Einstein
violation. A particularly instructive and useful rationalization of
this phenomenon was obtained studying Kinetically Constrained Models
(KCM)~\cite{FA84,GC02,BG03,BBL05,BCG05,JGC05,JSM08,JKGC08,toninelli}. These
are lattice models where a mobility field with local update rules
subjected to kinetic constraints reproduces the sluggish and
heterogeneous dynamics of glasses. The heterogeneous nature of
space-time correlations is explained in KCMs in terms of
\emph{defects} dynamics.  The activity field of KCMs is characterized
by rare \emph{mobility} defects which, wandering around in the system,
trigger the relaxation of the whole sample.  We show here that this
very same mechanism also leads to the anomalous transport properties
observed in~\cite{WHVB12,SH13}. Looking at the motion of a driven
intruder, the heterogeneous nature of the medium becomes manifest only
in the \emph{out-of-equilibrium} fluctuations $\sigma_x^2(t)$: this is
the case also for the diffusion on hierarchical lattices discussed
in~\cite{FCV14}, and for the field-induced superdiffusion of tracer in
a crowded medium discussed in~\cite{BBCILMOR13}.

We study the driven dynamics of a tracer particle in two kinetically
constrained models: the one-dimensional Fredrickson-Andersen
(FA)~\cite{FA84,GC02, BG03} and Bertin-Bouchaud-Lequeux (BBL)
models~\cite{BBL05,JSM08}.  Both have been studied and used as models
of supercooled liquids.  The local structure in FA is described by a
binary variable $n_i\in\{0,1\}$: sites with $n_i=1$ are active while
those with $n_i=0$ are inactive. The update of $n_i$ is possible only
when at least one among its neighbours is already active, namely one
needs $n_{i+1}=1$ or $n_{i-1}=1$. When possible, the update
$1\rightarrow 0$ is always accepted, while $0\rightarrow 1$ takes
place with probability $e^{-\beta}$, where $\beta$ is a dimensionless
inverse temperature. The dynamics obeys detailed balance, with an
energy function $E=\sum_i n_i$, so that the equilibrium state has no
correlations between sites. The FA model exhibits non-trivial
correlated dynamics for $\beta \gtrsim 1$. The concentration of active
sites is $c=\langle n_{i}\rangle = [1+ e^\beta]^{-1}$. Lengths are in
units of the lattice spacing which we set to one.  The BBL model is
described by a continuous variable: the density of mass $\rho_i\in [0,
  \infty)$.  The elementary step of the BBL dynamics is the
  simultaneous update of the density in a couple of neighbouring sites
  $\rho_i$ and $\rho_j$: $\rho_i'=q(\rho_i+\rho_j)$ and
  $\rho_j'=(1-q)(\rho_i+\rho_j)$, with $q\in [0,1]$. The update is
  possible only when the densities of the two sites fulfill the
  constraint $(\rho_i+\rho_j)/2< \rho_{th}=1$: the BBL is a
  kinetically constrained mass transport model. The random variable
  $q$, which introduces stochasticity in the dynamics of the density
  field, is extracted from a distribution $\psi_\mu(q)$ characterized
  by the parameter $\mu$. In the present study we consider the value
  $\mu=0.3$, which produces a diffusive dynamics of the mobility
  defects, which are represented by the \emph{active} links where the
  kinetic constraint is fulfilled. Details on the dynamics of the
  mobility defects for different values of $\mu$ can be found
  in~\cite{BBL05,JSM08}. In the BBL model the definition of
  \emph{active} links is naturally encapsulated into the definition of
  the model. In the FA model a link between two sites $i$ and $i+1$ is
  \emph{active} when both are active,
  $n_i=n_{i+1}=1$. \\ Following~\cite{BCG05,JGC05} we model
  micro-rheological experiments by assuming that the driven tracer can
  only move between two adjacent sites when these form an active link.
  Since the updates of $n_i$ in FA and of the mass field $\rho_i$ in
  BBL do not depend on the position of the probe the latter behaves as
  a~\emph{ghost} particle: it is influenced by the background but has
  no influence on it.  In order to induce a directed motion of the
  probe we unbalance in both models the probability of its forward and
  backward displacements: $p^{\rightarrow}_\epsilon = 1/2 +\epsilon$,
  $p^{\leftarrow}_\epsilon= 1/2 -\epsilon$, with $\epsilon\in
  [-1/2,1/2]$. The dynamics of a ghost probe in both the FA and the
  BBL can be then fully understood in terms of the mobility defects
  dynamics~\cite{GC02,BCG05}. The dynamics of the probe is ruled by
  two relevant time-scales: the mean \emph{persistence} time, which is
  the time the probe has to wait on average before being hit for the
  first time by a defect, and the average \emph{exchange} time, which
  is the time between two successive meetings with a defect. The
  difference between these two time-scales is both the signature of
  heterogeneous dynamics and the key ingredient of the anomalous
  transport of a probe. There is only one difference for the defects
  dynamics of the two models: while in the BBL the diffusion
  coefficient of defects does not depend on their concentration $c$,
  in the FA model the diffusion of mobile defects depends on the
  temperature, and hence on their concentration. Henceforth, in order
  to present a unified discussion for the two models, time is measured
  in the units of $\tau_0$, which is the time-scale on which a mobile
  defect moves of one step ($1/c$ and $1$ for the FA and the BBL
  models respectively). 
%%%% 
\begin{figure}
\includegraphics[width=\columnwidth]{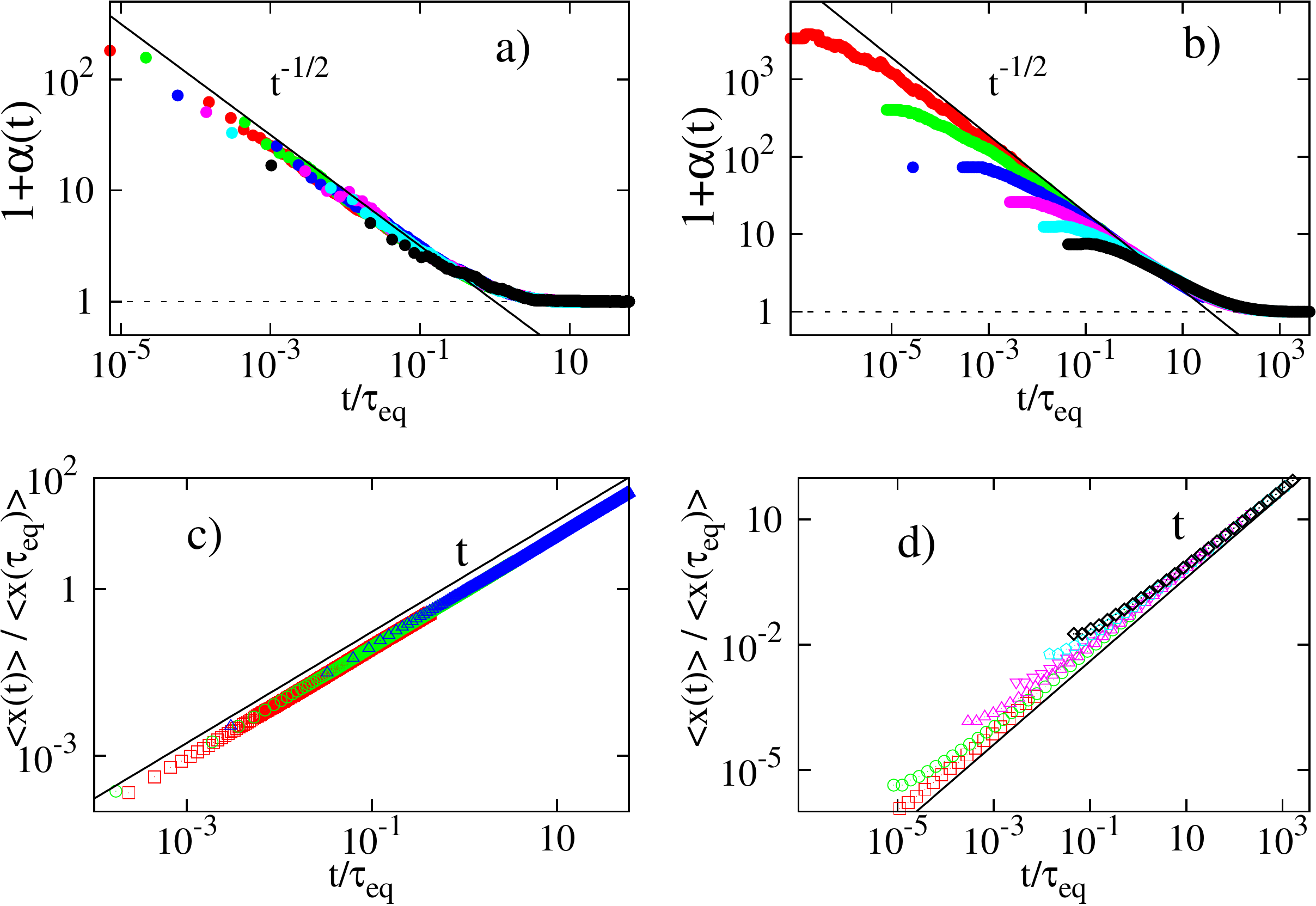}
\caption{ \emph{Panels a),b)}: Data collapse for the non-Gaussian
  parameter $\alpha_\epsilon(t)$ obtained plotting
  $1+\alpha_\epsilon(t)$ vs $t/\tau_{\textrm{eq}}$ for the different
  concentrations of mobility defects (different symbols) in FA [a)]
  and BBL [b)]. The continuous straight line emphasizes the behaviour
  $\alpha_\epsilon(t) \sim t^{-1/2}$ in the pre-asymptotic regime. The
  different concentrations of mobility defects are: $c=10^{-2}, 6.7
  \cdot 10^{-3}, 5.2 \cdot 10^{-3}, 3.8\cdot 10^{-3}, 2.8\cdot
  10^{-3}, 1.9\cdot 10^{-3}$ for the FA model; $c=2\cdot 10^{-1},
  1.1\cdot 10^{-1}, 5.1\cdot 10^{-2}, 1.6\cdot 10^{-2}, 2.7\cdot
  10^{-3}, 2.8\cdot 10^{-4}$ for the BBL model.  \emph{Panels c),d)}:
  Drift of the probe in FA [c)] and BBL [d)]: data are at the same
  concentrations of mobility defects of panels a) and b). Collapse is
  obtained plotting $\langle x_\epsilon(t)\rangle/ \langle
  x_\epsilon(\tau_{\textrm{eq}})\rangle$ vs $t/\tau_{\textrm{eq}}$.}
\label{fig1}
\end{figure} 
%%%

Assuming that defects are independent random walkers the dynamics of
the probe can be described as a Continuous Time Random Walk
(CTRW)~\cite{BCG05}. In this approximation the histogram of probe
displacements, which corresponds to the self part of the van Hove
function $G_s(r,t)=\langle \delta(r-[x(t)-x(0)])\rangle$ (the angular
brackets indicate the average over different trajectories), can be
exactly written with the following formula~\cite{BCG05}: \be
G_s(x,t)=\mathcal{P}(t) \delta(x) + \int_0^t p(t-s)
P_{\textrm{1st}}(x,s) ~ds. \label{PofX} \ee In Eq.~(\ref{PofX}) the
probability of persistence times is denoted by $p(t-s)$,
$P_{\textrm{1st}}(x,s)$ is the propagator for trajectories which start
with a jump event (the subscript $_{\textrm{1st}}$ indicates that at
least one displacement took place) and $\mathcal{P}(t)$ is the
persistence function, i.e., the probability that a probe is not hit by
a mobility defect for a duration of $t$ when the observation starts at
an arbitrary time. From Eq.~(\ref{PofX}) is clear that the population
(or the probability) is split into two groups: probes which at time
$t$ have already started to move and probes which at $t$ are still at
rest (we refer to a population of probes since we can think of having
many probes evolving in parallel and starting from random positions).
Since in both FA and BBL models defects behave as independent random
walkers, persistence equals the survival probability $\mathcal{P}(t)=
e^{-c \sqrt{t}}$, with $c$ the concentration of
walkers~\cite{RBook}. The distribution of persistence times $p(t)$
which enters Eq.~(\ref{PofX}), and is obtained as
$p(t)=-d\mathcal{P}(t)/dt=c~e^{-c \sqrt{t}}/\sqrt{t}$. The
distribution of exchange times, which is in turn proportional to
$-dp(t)/dt$~\cite{JGC05}, reads to leading order in $c$ as
$\psi(\tau)=c~e^{-c \sqrt{\tau}}/\tau^{3/2}$. The representation of
the probe dynamics as a CTRW is very insightful and will be our main
tool to understand the relationship between dynamic heterogeneity and
anomalous diffusion of a driven tracer.

\begin{figure}
\includegraphics[width=\columnwidth]{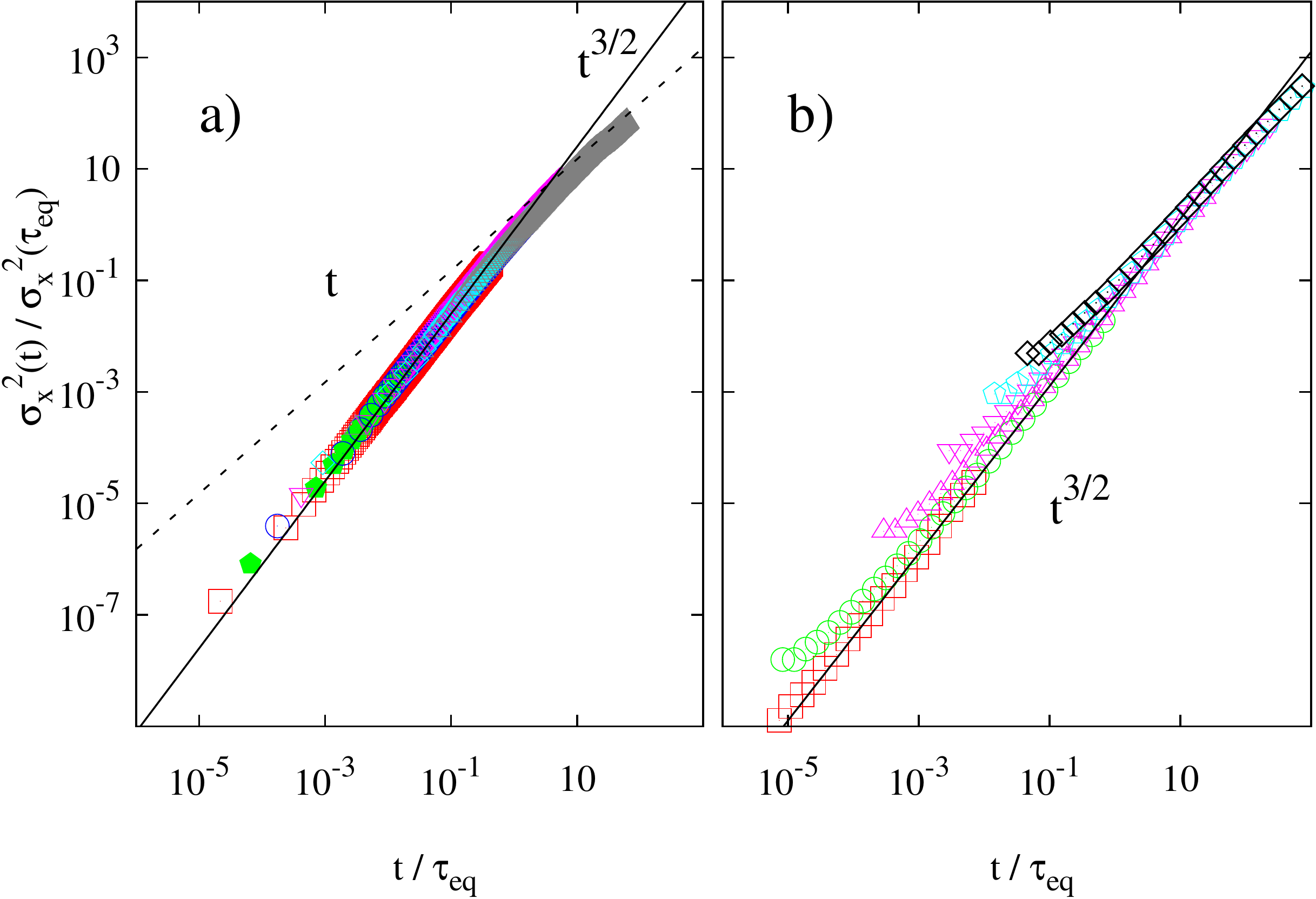}
\caption{Mean square displacement around the drift $\sigma_x^2(t)$ for
  a driven probe ($\epsilon=1/2$) in FA [panel a)] and BBL [panel
    b)]. Different symbols represent different concentrations $c$ of
  the mobility defects, which are, respectively for the two models,
  the same as in Fig.~\ref{fig1}. Collapse of the curves is obtained
  by plotting $\sigma_x^2(t)/\sigma_x^2(\tau_{\textrm{eq}})$ vs
  $t/\tau_{\textrm{eq}}$. Full line is the $t^{3/2}$ scaling, dashed
  line [only panel a), FA] is the Fickian behaviour $\sigma_x^2(t)\sim
  t$.}
\label{fig2}
\end{figure}

We now present our numerical results about anomalous diffusion in the
FA and BBL models, which are shown in Fig.~\ref{fig1} and
Fig.~\ref{fig2}. By looking at Fig.~\ref{fig1} we notice that a driven
probe ($\epsilon=1/2$) has a linear drift in both FA and BBL, but at
the same time the non-Gaussian parameter $\alpha_\epsilon(t)$ for
non-centered distributions signals important deviations from
Gaussianity up to $t\sim \tau_{\textrm{eq}}$. For a Gaussian
distribution with mean $\langle x_\epsilon(t)\rangle$ and variance
$\sigma_x(t)$ the fourth non-centered moment reads $\langle
x_\epsilon^4(t) \rangle = \langle x_\epsilon(t)\rangle^4+ 6 \langle
x_\epsilon(t)\rangle^2\sigma_x^2(t)+3\sigma_x^4(t)$, which allows us
to define $\alpha_\epsilon(t)$ as: \be \alpha_\epsilon(t) =
\frac{\langle x_\epsilon(t)^4 \rangle}{\langle x_\epsilon(t)\rangle^4
  + 6 \langle x_\epsilon(t)\rangle^2\sigma^2_x(t)+3\sigma^4_x(t)}-1,
\ee where $\sigma^2_x(t)=\langle x^2_\epsilon(t)\rangle-\langle
x_\epsilon(t)\rangle^2$. It can be easily seen that with zero drift
the standard definition of the non-Gaussian parameter is
recovered. For aesthetic reasons in Fig.~\ref{fig1} we plotted
$1+\alpha_\epsilon(t)$.  Eq.~(\ref{PofX}) tells us that the overall
drift comes from the convolution of the drift of the moving probes
with the distribution of persistence times: \be \langle x_\epsilon(t)
\rangle = \int_0^t~ds~p(t-s) \langle
x_\epsilon(s)\rangle_{\textrm{1st}}. \label{moments} \ee From the
inspection of Eq.~(\ref{moments}) is possible to single out the
different physical mechanisms~\cite{IBOV14} which determine the linear
behaviour of the drift, $\langle x_\epsilon(t) \rangle \sim t$, in the
two regimes $1\ll t\ll \tau_{\textrm{eq}}$ and $\tau_{\textrm{eq}}\ll
t$. In the latter, due to the exponential cut-off of $\psi(\tau)$, for
$t\gg \tau_{\textrm{eq}}= c^{-2}$ the drift $\langle x_\epsilon(s)
\rangle_{\textrm{1st}}$ is linear. In this regime one can approximate
$\int_0^t~ds~p(t-s) \langle x_\epsilon(s) \rangle_{\textrm{1st}} \sim
(1-\mathcal{P}(t)) \langle x_\epsilon(t) \rangle_{\textrm{1st}} \sim
\langle x_\epsilon(t) \rangle_{\textrm{1st}} $, which shows that the
total drift is also linear. On the contrary, in the former regime that
we will call pre-asymptotic henceforth, persistence and exchange
distributions can be approximated by power-law distributions:
$\psi(\tau)\sim \tau^{-3/2}$ and $p(t)\sim t^{-1/2}$.  This leads to a
subdiffusion of the moving probe as $\langle x_\epsilon(t)
\rangle_{\textrm{1st}} \sim \sqrt{t}$. The physical reason is that
moving probes are repeatedly hit by a mobile defect $t^{1/2}$
times. On the other hand the fraction of moving probes increases too,
as $ c~\sqrt{t}$, due to the heavy tail of $p(t)$. How these two
effects combine can be read off in the explicit expression of
Eq.~(\ref{moments}) in the regime $t\ll \tau_{\textrm{eq}}$, i.e.,
$\langle x_\epsilon(s) \rangle \sim c~\int_0^t
ds~(t-s)^{-1/2}~\sqrt{s}$: the change of variable $s\rightarrow s/t$
in the last integral yields immediately $\langle x_\epsilon(s) \rangle
\sim c~t$. It is due to this non-trivial mechanism that, even in the
pre-asymptotic regime, we can observe a linear drift.

Fig.~\ref{fig2} shows then that the non-Gaussianity of
$G_\epsilon(x,t)$ is manifest in the transport properties of the
intruder when one looks at the MSD around the drift
$\sigma_x^2(t)=\langle x_\epsilon^2(t)\rangle - \langle x_\epsilon(t)
\rangle^2$: it grows superdiffusively as $\sigma_x^2(t) \sim t^\gamma$
with $\gamma \approx 3/2$ for $1\ll t\ll \tau_{\textrm{eq}} $.  The exponent
$3/2$ is the same which characterizes the field-induced superdiffusion
of a tracer in a crowded medium~\cite{BBCILMOR13}.
%%%
The population splitting scenario, recently discussed also in the
context of CTRW with aging dynamics~\cite{SBM13,SBM14}, allows one to
perfectly understand both qualitatively and quantitatively not only
the observed superdiffusion (Fig.~\ref{fig2}), but also the
pre-asymptotic behaviour of the non-Gaussian factor (Fig.~\ref{fig1}).

From the definition of $G_s(x,t)$ in Eq.~(\ref{PofX}) we have that the
MSD around the drift reads \be \sigma_x^2(t) =
-\langle x_\epsilon(t)\rangle^2 + \int_0^t
ds~p(t-s)~\langle x_\epsilon^2(s) \rangle_{\textrm{1st}}.
\label{MSDpers}
\ee We already know that in the pre-asymptotic regime $\langle
x_\epsilon(t)\rangle\sim c~t$, so that we only need to know $\langle
x_\epsilon^2(s) \rangle_{\textrm{1st}}$. The latter is obtained from
the asymptotic behaviour of moments for a biased
CTRW~\cite{CMMV99,GSVV12,BGSVV13,BGSVV14}.  In a CTRW with waiting
time distribution $\psi(\tau)\sim\tau^{-(1+\beta)}$ and $0<\beta<1$ it
holds $\langle x_\epsilon^2(t)\rangle \sim t^{2\beta}$: since in our
case $\beta=1/2$, we have $\langle
x_\epsilon^2(s)\rangle_{\textrm{1st}} \sim s$. By plugging this last
result in Eq.~(\ref{MSDpers}) and retaining the leading contribution
to $p(t-s)$ when $t\ll \tau_{\textrm{eq}}$, one finds \be
\sigma_x^2(t) \sim c~\int_0^t ds~\frac{s}{\sqrt{t-s}}\sim
c~t^{3/2}, \label{MSDanom} \ee where the asymptotic behaviour of the
integral has been evaluated by simply changing variable $s \rightarrow
s/t$. The term $\langle x_\epsilon(t)\rangle^2 =O(c^2)$ appearing in
Eq.~(\ref{MSDpers}) has been dropped in Eq.~(\ref{MSDanom}) because it
is subleading in the regime $c \ll 1$, $t\ll \tau_{\textrm{eq}}$. We
obtained analytically the superdiffusive behaviour tracing it back to
the subsequent hits with the mobile defects, which are in turn encoded
in the heavy tails of $\psi(\tau)$. To estimate the pre-asymptotic
behaviour of the non-Gaussian parameter $\alpha_\epsilon(t)$ we need
to know the fourth order non-centered moment, which from
Eq.~(\ref{PofX}) reads: \be \langle x^4_\epsilon(t) \rangle = \int_0^t
ds~ \frac{c}{\sqrt{t-s}}~\langle x^4_\epsilon(s)
\rangle_{\textrm{1st}} \sim c~t^{5/2}.
\label{x4}
\ee The scaling in Eq.~(\ref{x4}) comes again from the asymptotic
behaviour of moments in a CTRW characterized by the distribution of
waiting times $\psi(\tau)\sim\tau^{-(1+\beta)}$ and $0<\beta<1$:
$\langle x^4_\epsilon(s) \rangle \sim
s^{4\beta}$~\cite{BGSVV13,BGSVV14}. Since in our case $\beta=1/2$ we
have $\langle x^4_\epsilon(s) \rangle \sim s^2$. For small values of
$c$ the denominator of the non-Gaussian parameter reads at leading
order: \be \langle x_\epsilon(t) \rangle^4 + 6 \langle x_\epsilon(t)
\rangle^2 \sigma_x^2(t) + 3 \sigma_x^4(t) \sim 3~c^2~ t^3, \label{den}
\ee so that, combining Eq.~(\ref{x4}) with Eq.~(\ref{den}) we find:
\be \alpha_\epsilon(t) \sim c^{-1}t^{-1/2} \label{nGauss}, \ee which
is perfectly consistent with the numerical behaviour of
$\alpha_\epsilon(t)$ shown in Fig.~\ref{fig1}. \emph{Strong} anomalous
diffusion takes place when the scaling assumption
$G_s(x,t)=\mathcal{F}[x/\ell(t)]/\ell(t)$ for the van Hove function
and the scaling of moments, $\langle x^n(t)\rangle \sim \ell^n(t)$,
\emph{cannot} be written in terms of a single length
$\ell(t)$~\cite{CMMV99}. In the present case (FA and BBL) this
phenomenon takes place due to population splitting. Looking at the
distribution of moving probes we have found the numerical evidence
(not shown) that $P_{\textrm{1st}}(x_\epsilon,t)$ is a half-Gaussian:
at different times a perfect collapse of data is obtained with
$\ell(t)=t^{1/2}$. Taking otherwise into account the whole population
of probes, i.e. also the contribution $\mathcal{P}(t)\delta(x)$, in
Eq.(\ref{PofX}), of probes which never jumped up to time $t$, we have
$\sigma_x^2(t) \sim t^{3/2} \neq \ell^2(t) = t$: diffusion is
\emph{strongly} anomalous~\cite{CMMV99}. Let us stress that this
strong anomalous diffusion is not due, as usual, to a multiscaling
property of the probability distribution of
displacements~\cite{CMMV99}: it comes from the splitting of probes
population into slow persistent ones, not moving roughly until $t \sim
\tau_{\textrm{eq}}$, and those which already at $t \ll
\tau_{\textrm{eq}}$ have already been repeatedly hit by mobility
defects. Precisely the same mechanisms is at the origin of S-E
violation $D~\tau_{\textrm{eq}}=e^{-\beta}$: a probe diffuses much
more than one step on the relaxation time-scale due to repeated
interactions with the same mobile defect. Such a mechanism is also
responsible for dynamic heterogeneity: all regions that relax because
they are hit by the same mobile defect within the time-scale
$\tau_{\textrm{eq}}$ become dynamically correlated. We have therefore
shown that dynamic correlations, violation of S-E and \emph{strong}
anomalous superdiffusion are directly connected in KCMs \cite{BCG05}
(see also \cite{toninelli}).\\

In conclusion we have related the superdiffusive behaviour of driven
probes to the splitting of their population in frozen ones and moving
ones repeatedly hit by the same mobility defect. Our analysis provides
an explanation of the results found in atomistic
models~\cite{WHVB12,SH13} and link them to the phenomena of dynamic
heterogeneity and Stokes-Einstein violation. Furthermore, it offers a
theoretical derivation of the superdiffusion exponent $3/2$. This
value, compatible with our numerical results for both the FA and the
BBL model, is also surprisingly close to the value $1.45$ found in the
supercooled Yukawa mixture of~\cite{WHVB12}.  It is also remarkable
the qualitative agreement between the behaviour of the non-Gaussian
factor $\alpha_\epsilon(t)$ characterizing the driven pre-asymptotic
dynamics of a probe in one-dimensional FA and BBL models and the one
found for the unbiased dynamics of a probe in a three-dimensional
supercooled Lennard-Jones mixture (see Fig. 3 of~\cite{SH13}). All
these similarities point towards the presence of universal features,
still to be investigated, which are intrinsically related to dynamical
heterogeneity and emerge at low temperatures independently from the
dimensionality and the specific interactions of the models. We expect
indeed our findings to hold generically beyond the simple models we
focused on.  In fact, the key ingredients are the population splitting
scenario and the anomalous diffusion of mobile probes induced by
dynamic correlations, which are phenomena known to be present
generically in supercooled liquids.

\begin{acknowledgments}
G.G. acknowledges support from the ERC Grant GLASSDEF No.~ADG20110209.
G.B. acknowledges support from the ERC Grant NPRGGLASS.  We wish to
thank E. Barkai, J.-L. Barrat, O. B\'enichou and A. Heuer for useful discussions
and comments.
\end{acknowledgments}

%\bibliography{apssamp}% Produces the bibliography via BibTeX.

\begin{thebibliography}{61}
\bibitem{foot1} One can consider all the extra factors as constant since their variation is very slow.
\end{thebibliography}

\end{document}